\documentstyle[epsfig,natbib2,natbibmnfix]{mn2e}

\newcommand{\be}{\begin{equation}}
\newcommand{\ee}{\end{equation}}

\newcommand{\apj}{ApJ}
\newcommand{\apjs}{ApJS}
\newcommand{\mnras}{MNRAS}
\newcommand{\aap}{A\&A}
\newcommand{\araa}{ARA\&A}
\newcommand{\apjl}{ApJL}
\newcommand{\aj}{AJ}
\newcommand{\nat}{Nature}

\def\ltsima{$\; \buildrel < \over \sim \;$}
\def\simlt{\lower.5ex\hbox{\ltsima}}
\def\gtsima{$\; \buildrel > \over \sim \;$}
\def\simgt{\lower.5ex\hbox{\gtsima}}
\def\sgra{Sgr~A$^*$}

\def\msun{{\,{\rm M}_\odot}}

\def\lsun{{\,L_\odot}}

\def\del#1{{}}

\title[]{The ``missing"  YSOs in the central parsec of the Galaxy:
evidence for star formation in a massive accretion disk and a top-heavy IMF}

\author[S.~Nayakshin and R. Sunyaev]
{\parbox{18cm}{Sergei Nayakshin$^{1,2}$ and Rashid Sunyaev$^{2}$}\vspace{0.3cm}\\
$^1$Dept. of Physics \& Astronomy, University of Leicester, Leicester, LE1 7RH, UK\\
$^2$Max-Planck-Institut f\"{u}r Astrophysik, Karl-Schwarzschild-Stra\ss{}e 1,
85741 Garching bei M\"{u}nchen, Germany}

\begin{document}

\maketitle

\begin{abstract}
Few dozens of young high mass stars orbit \sgra\ at distances as short as $\sim 0.1$ parsec, 
where star formation should be quenched by the strong tidal shear from \sgra. The puzzling young stellar population is believed to come into existence in one of the two ways:
 (i) ``normal" star formation at several tens of parsec in a very massive star cluster that then spiraled in, or (ii) star formation in situ in a massive self-gravitating disk. We propose to constrain these two 
scenarios via the expected X-ray emission from young low mass stars that should have formed alongside the massive stars. To this end we compare the recent {\em Chandra} observations of X-ray emission from young stars in the Orion Nebula, and the {\em Chandra} observations of \sgra\ field. We show that the cluster inspiral model is ruled out  irrespectively of the initial mass function (IMF) of the young stars. In addition, for the in situ model, we find that no more than few thousand low-mass stars could have formed alongside the massive stars. This is more than a factor of ten fewer than expected {\em if} these stars were formed with the standard IMF as elsewhere in the Galaxy. The young stars in the GC are thus the first solid observational evidence for star formation in AGN disks and also require the IMF of these stars to be top-heavy. We briefly consider implication of these results for AGN in general.
\end{abstract}

\renewcommand{\thefootnote}{\fnsymbol{footnote}}
\footnotetext[1]{E-mail: {\tt Sergei.Nayakshin {\em at} astro.le.ac.uk}}

\section{Introduction}
\label{intro}

\sgra\ is the closest super-massive black hole (SMBH), naturally residing in the centre of our Galaxy \citep[e.g., ][]{Schoedel02,Ghez03b}, and weighing about $M_{\rm BH} \sim 3.5 \times 10^6 \msun$. Few tens of close young massive "He-I" emission line stars dominate the energy output of the central parsec of the Galaxy \citep{Krabbe95}. Almost all of these stars belong to one of two {\em stellar} rings \citep{Levin03,Genzel03a}. The ages of the young stars are estimated at $t \approx 4\pm 1$ Million years (Genzel 2005, private communication), and the initial masses of the He-I stars are probably larger than $M\simgt 40 \msun$ \citep{Krabbe95,Paumard01}.   ``Standard'' modes of star formation at $R= 0.1R_{0.1}$ pc distances from a SMBH are forbidden due to a huge tidal field of the central object as the required gas density is $n_H 
\simgt 10^{11} \hbox{cm}^{-3}  R_{0.1}^{-3}$. Understanding formation mechanisms of these stars is a challenge and a test for both star formation theory and AGN phenomenon as such.

Two models suggested to
resolve this difficulty do so in two opposite ways. \cite{Gerhard01} showed that the stars may have been formed at distance of tens of parsecs, thus avoiding the need for excessive gas density prior to star formation, in a massive star cluster. The star cluster's orbit would then decay through dynamical friction with the background stars. Further detailed models however showed that the cluster needs to be very massive ($M \sim 10^6\msun$) and very compact 
\citep{Kim03,Hansen03,McMillan03,Kim04,Gurkan05}.

The other model \citep{Levin03,Milosavljevic04,NC05} overcomes the tidal density limit by sheer force (mass), which naturally occurs at large distances in a 
standard accretion disk model \citep{Shakura73} when the disk mass exceeds a fraction of a percent or so of the SMBH mass \citep{Paczynski78,Kolykhalov80,Collin99,Goodman03}. 
Stars may then be formed directly in situ 
{\em if} radiative cooling is efficient enough \citep{Shlosman90,Gammie01}. 

These two scenarios of young star formation could perhaps be differentiated if we knew the stellar content of the young disks better. Indeed, the accretion disk may become gravitationally unstable when the gas mass is $M \simgt 10^4 \msun$ \citep{NC05}, whereas for the infalling star cluster model at least $10^6\msun$ material is needed, and it is thus likely that the former model would produce much less of the stars than the latter. In addition, irrespectively of the two models proposed for the He-I stars, the observational knowledge of IMF in \sgra\ could shed light on whether the IMF is a universal function or is a function of environment. The problem one faces when attempting to constrain the number of young stars in the infrared observations of the GC, however, is that a million of a solar-type stars emit bolometrically as little as a single bright O-star. In particular, \cite{Meyer05} calculates that low-mass pre-main sequence stars may account for (as much as or as little as) $\sim 4-12$\% of the integrated light at the near-infrared frequency for young star bursts, and suggests that high resolution observations may discern that in high resolution spectra. It is not clear to us if this method could be applied to the extremely bright and confusion-limited field of the GC  \citep{Genzel03a,Ghez05}. 

Here we point out that young (pre-main sequence; PMS hereafter) low mass stars are
strong X-ray emitters (see \S \ref{sec:orion}), and that due to the extremely low X-ray luminosity of \sgra\ field it is possible to set interesting limits on the population of these stars there. We find that, due to the very high initial stellar mass, the cluster infall model is ruled out by these limits independently of whether the IMF is steep or top-heavy. The in situ star formation model remains viable if formation of low mass stars was suppressed  by at least a factor of ten compared with the normal galactic IMF. We also note that, in contrast to \sgra\ star cluster, the two other massive young star clusters near the GC appear to have more or less normal IMF, indicating that \sgra, the SMBH, is the  decisive factor in forming the top-heavy IMF in the stellar disks around it.


\section{Orion Nebula YSO X-ray emission}\label{sec:orion}

The Orion Nebula (ON) is one of the closest sites of recent star formation which has been well studied in the optical and near infrared, with IMF that appears to be very much consistent \citep{Hillenbrand97}  with the standard galactic IMF  \citep[e.g.,][]{Miller79}.  Recently, the 
{\em Chandra} Orion Ultradeep Project (COUP) has culminated in an unprecedentedly large and detailed catalog of X-ray properties of Young Stellar Objects (YSO)\footnote{Following the review by \cite{Feigelson99}, we shall refer to all low to intermediate PMS stars as YSOs}  in the Orion Nebula in the range of ages between less than a Million to somewhat less than $10$ Million years.  The combined multi-waveband studies of YSO in Orion result in a consistent picture of YSO's X-ray properties. 

Globally, the $\sim 1400$ identified low mass ON stars ($M \simlt 3 \msun$; the higher intermediate mass stars are weak in X-rays) emit in the hard\footnote{YSOs also emit an about equal luminosity in softer X-ray band, but these X-rays would be mainly absorbed in the large column of neutral material on the line of sight to the GC, $N_H\simeq 10^{23}$ H atoms/cm$^2$ \citep{Baganoff03a}.} 
X-ray band (2-8 keV) the total of  $L_{\rm ON}  = 1.2 \times 10^{33}$ erg/sec \citep{Feigelson05a}. 
Stellar X-ray luminosities show a very slow decay in the first $\sim 10^7$ years, i.e. during the PMS stage of stellar evolution,  with $L_X \propto t^{-1/3}$ \citep{Preibisch05}. The X-ray emission is understood to result mainly from magnetic flares on the surface of the stars rather than accretion, which is evidenced by (i) rotational modulation of X-ray emission \citep{Flaccomio05} with stellar rotation period; (ii) a decrease in the average X-ray luminosity of the stars when accretion disks are present \citep{Kastner05}; (iii) the presence of frequent X-ray flares \citep[e.g.,][]{Glassgold05}. Also note that accretion disks around YSOs disappear much sooner \citep[e.g.][]{Hernandez05} than does the X-ray emission from YSOs \citep{Preibisch05}. 
The X-ray emission of YSOs should thus be related to the very way the young low mass stars relax to their main-sequence structure rather than be a result of YSO's interactions with gas disks.
The total mass of the identified 1600 young stars in the ON is around $M_{\rm ON} \simeq 1000 \msun$ \citep{Hillenbrand97} .  \cite{Hillenbrand97} notes that ON stars more massive than $10 \msun$ weigh in total $160 \msun$, or about 0.085 of the total (identified plus unidentified sources) cluster mass, exactly as predicted by the \cite{Miller79} IMF. 

\section{X-ray emission from YSO population in \sgra}

\subsection{In situ star formation}\label{sec:insitu}

We first estimate the X-ray emission expected from the young stars in \sgra\ young star cluster.
We now assume that \sgra\  young stars, and for definitiveness we only consider those in the larger and better determined clock-wise stellar disk \citep{Levin03,Genzel03a}, have IMF close to that of \cite{Miller79}, as the ON stellar cluster does \citep{Hillenbrand97}. The "He-I emission line stars" have estimated birth masses of larger than $M_*\simgt 40 \msun$ \citep{Paumard01}.  With $\simeq 10$ of such stars in the clock-wise system, and with stars on average a bit more massive than the estimated minimum mass above, the total minimum initial disk mass for stars with $M_* > 40 \msun$ is around $500 \msun$. Extrapolating this to $10 \msun$ we expect to have $M_{10}\sim 3000 \msun$ in stars more massive than $10 \msun$  in the clock-wise disk, initially, which is about 20 times more than that in the ON cluster.  Then, quite simply, the total expected X-ray luminosity of young stars less massive than $\sim 3 \msun$  in \sgra\ cluster is 20 times that of the YSO hard emission in the ON. The expected stellar disk luminosity in the photon energy range $2-8$ keV is thus
\begin{equation}
L_{\rm exp} \approx 2.5 \times 10^{34}  \; \frac{M_{10}}{3000}\; \quad \hbox{erg/sec}\;.
\end{equation}
Note that our normal IMF estimate of the total mass of the \sgra\ young star cluster, $M_{\rm total} \sim 2 \times 10^4 \msun$, agrees well with other previous estimates. For example, \cite{Figer04} estimates the total mass of \sgra\ young star cluster as $M_{\rm total} = 10^4 \msun$ by assuming a \cite{Salpeter55} IMF down to $M_* = 1 \msun$. Since we have assumed an IMF that extends to lower $M_*$, such as in the ON cluster, we obtained a slightly higher estimate.

Now, we can compare this to the diffuse X-ray emission observed inside the projected area of the clock-wise disk. This system has a size of about $R_{\rm disk}=5"$, and is inclined to the line of sight at $i=126^\circ$ \citep{Genzel03a,Paumard05}. the projected area is thus $A=\pi R^2 \cos i = 46$ arcsec$^2$. Further, \cite{Baganoff03a} find that the diffuse X-ray surface brightness in the inner 10 arcsecond of the Galaxy is $dL_x/dA = 7.6\times 10^{31}$ erg/sec/arcsec$^2$. However the spectrum of the diffuse background emission can be fit by a thermal plasma with temperature of only $k T = 1.3$ keV (see \S 7 in Baganoff et al. 2003), whereas the hard component of the Orion YSO spectra can be fit by a thermal spectrum with $kT \simeq 3$ keV. The harder YSO spectrum would thus be well discernible in the observed spectrum of the \sgra\ diffuse emission had the former been there. We therefore estimate that YSO conribution to the diffuse emission in the projected area of the young clock-wise stellar disk can be no more than a third of the observed X-ray luminosity,
\begin{equation}
L_{\rm stellar} < \frac{1}{3}\; \frac{dL_x}{dA}\; A = 1.2 \times 10^{33} \frac{R_{\rm disk}^2}{(5")^2} \quad \hbox{erg/sec}\;.
\end{equation}

We see that there is a disparity of a factor of 20 or so between the expected total X-ray luminosity of young low mass stars and what could be attributed to their emission in the actual observations of \sgra.  Thus, one of our assumptions, e.g., that the stars were formed in situ, or that they had a normal galactic IMF, needs to be abandoned.

\subsection{Infall of a massive star cluster}

 By design \citep{Gerhard01}, the young cluster starts off at a larger initial radius than the current location of the massive young stars, e.g. at distance $R = R_{\rm cl} $ parsec. As it spirals in, the low mass stars are peeled off the cluster first, whereas the high mass stars, concentrated in the cluster's center, make much further in. State of the art Monte Carlo simulations by \cite{Gurkan05} show that in order to transfer the massive stars into the inner 0.1 parsec, the cluster needs to be as massive as $M_{\rm cl}  > 10^6 \msun$. This is about three orders of magnitude more massive than the identified stars in the ON star cluster. The expected luminosity of the low mass YSOs from the inspiralling cluster can then be estimated as a factor of a 1000 that of the ON:
\begin{equation}
L_{\rm exp} \approx 10^{36}  \; \frac{M_{\rm cl}}{10^6 \msun}\; \quad \hbox{erg/sec}\;.
\end{equation}

Now, the observed diffuse X-ray emission \citep{Muno04}  inside a projected distance $R_{\rm cl}$ scales as
\begin{equation}
L_{\rm diff} \simeq 3.5 \times 10^{33} R_{\rm cl}^2 \quad \hbox{erg/sec}\;.
\end{equation}
where $R_{\rm cl}$ is in parsecs. Hence, only if the cluster initial position was $R_{\rm cl} \simgt 30$ would the YSO's be undetectable amongst the observed X-ray diffuse emission. None of the clusster cores in the simulations by \cite{Gurkan05} that start off at similarly large radii reach the sub-parsec region of the GC, however. In addition, as pointed out by \cite{Kim03} and \cite{Gurkan05}, the standard galactic IMF would then predict hundreds to thousands of He-I stars rather than dozens observed, and most of them would be peeled off the cluster outside the central parsec, which would strongly contradict the observations. 

If these arguments were not enough, there is also the fact that YSOs produce X-ray flares, some of which reach luminosities of $L_x > 10^{32} $ erg/sec \citep{Favata05}, which would be observable by {\em Chandra} in the relatively low surface brightness area outside the inner parsec. In fact, this argument has already been given by \cite{Muno04} who pointed out that only 5 such candidate flares were observed in the inner $R=20$ parsec of the Galaxy. These authors went on to note that since about $0.1$\% of YSO are flaring at the required levels at any given time, this implies the upper limit on the number of YSOs in this region is about 5000. This number is short by more than 2 orders of magnitude from the $\sim 10^6$ estimate based on cluster inspiral simulations. While the new more complete statistics from the COUP project could allow a more detailed estimate of the maximum YSOs number in the central $\sim 30$ parsecs than done by \cite{Muno04}, it is very unlikely that more than two orders of magnitude shortage of the YSOs could be accounted for in this way (Muno 2005, private communication).

\section{Discussion}

\subsection{Implications for \sgra}

Comparison of the expected emission of YSOs within  the mass range from a fraction of to a few Solar masses, to the observed emission in the central tens of parsec from \sgra\ puts tight limits on how many YSOs could have formed alongside the young He-I emission line stars near \sgra. The infalling young star cluster model appears to be ruled out by these limits, irrespectively of the assumptions made about the IMF of the cluster. Indeed, as shown above, with a "normal" steep IMF, there should be as many as $10^6$ YSOs. These are missing by a factor of a 100 or so in the inner 30 parsec. Further, introducing a top-heavy IMF where, say, 99\% of the mass is in the high mass stars ($M>10\msun$, for example) would resolve the problem with the YSOs X-ray emission, but would create a similarly severe problem of over-producing the high mass stars themselves. These stars would be scattered around a much larger region than the inner parsec \citep{Kim03}, in disagreement with the observations \citep{Genzel03a,Ghez05}.

The accretion disk star formation model fares better. This model can be reconciled with the data by requiring a top-heavy IMF where the low mass stars are suppressed by a factor of ten or so.  
We therefore feel that the massive stars near \sgra\  are the first observational evidence for a star formation in an accretion disk.  Similar but quantitatively less constraining conclusions were reached by Nayakshin et al. (2005, to be submitted) on a completely independent basis of stellar orbits for the young stars in the GC.

Several authors have previously suggested \citep[e.g.,][]{Morris93,Krabbe95,Levin03} that due to the extreme environment near the SMBH in our Galactic Centre, formation of low mass stars may be much less efficient than elsewhere in the Galaxy. Convincing quantitative arguments were never made, however, as any stellar mass estimates, such as a Jeans mass estimate, depend on a range of poorly known parameters (initial gas density, temperature, or the quantities determining these). A detailed theoretical analysis of the implications of the observational constraints found here will be given elsewhere.

\subsection{Comparison to other young star clusters in the GC region}

The Galactic Centre region "spans the central few hundred parsecs of the Galaxy and represents about 0.04\% of the volume in the Galactic disk. Although small in size, the region contains 10\% of the molecular material and young stars in the Galaxy..." \citep{Figer01}. In particular, the inner $\sim 50$ parsec of the Galaxy contain three massive young star clusters, the Arches, the Quintuplet and the Central (the \sgra\ star cluster). Their heavy mass content ($M \simgt 20 \msun$) is very similar, with each containing about 100 stars, but the Arches is $2-3$ million years old versus $3-6$ million years for the other two clusters \citep{Figer04}. In view of our results, it is interesting to compare the \sgra\ young star cluster to the other two star clusters. The best studied of the two is the Arches star cluster. The estimated mass of the cluster is $\sim 1.2 \times 10^4 \msun$ \citep[e.g.][]{Stolte02}, and its stellar density is derived to be of same order or larger than that of the young \sgra\ star cluster \citep[Table 1 in ][]{Figer04}. The IMF is much flatter than the classical \cite{Salpeter55} IMF in the innermost few arcseconds  of the cluster, but it steepens to an IMF steeper than the classical one at the outer regions of the cluster \citep{Stolte02}. This is interpreted as evidence for a dynamical mass segregation in which massive stars sink close to the center of the cluster \citep{Portegies-Zwart02}, rather than evidence for the IMF to be top-heavy originally. 

As a self-consistency check, it is also interesting to estimate the X-ray emission from young low mass stars in the Arches cluster and compare it to the observed spectrum in a way similar to what we did for \sgra. Repeating the arguments similar to those presented in \S \ref{sec:insitu}, we would estimate the expected YSOs hard X-ray emission at $\sim (1-2) \times 10^{34}$ erg/sec. 
\cite{Yusef-Zadeh02} find that the hard X-ray luminosity of the Arches cluster is about $L \sim 10^{35}$ erg/sec, \citep[probably dominated by colliding fast stellar winds, e.g.][]{Yusef-Zadeh02,Rockefeller05}.
Without a detailed analysis it is hard to say with a complete certainty that the expected YSO's X-ray emission can be hidden in the observed spectrum, but it does appear quite plausible to us.

For completeness of the discussion, we note that the dynamical mass segregation would not have worked in \sgra\ in contrast to the other two young star clusters. The He I stars, the most massive in the \sgra\ stellar system, belong to the two well defined stellar rings \citep{Levin03,Genzel03a,Paumard05}, that is two not relaxed systems.  On the theoretical side, while the stars near \sgra\ are a factor of $\sim 2$ older than stars in the Arches cluster, the former have a much higher velocity dispersion than the latter because of the SMBH presence. The \sgra\ star cluster relaxation time is thus much longer than the age of the young stars there, whereas this is not true for the Arches cluster \citep{Portegies-Zwart02}.

Therefore, \sgra\ is distinct from the other two GC young star clusters in its lack of low mass YSOs, and it seems obvious that it owes its uniqeness to the SMBH in the centre of the Galaxy.

\subsection{Implications for AGN in general}

If top-heavy IMF  is a general feature of star formation in AGN disks, then several immediate consequences result. First of all, stellar feedback is then more important  for accretion and the whole AGN phenomenon than thought based on a standard IMF. For example, \cite{Krolik88} showed that with the usual IMF, star formation inside the (putative) molecular torus \citep[e.g][]{Antonucci93} around AGN would fail to account for the large random speeds observed there, as the low mass young stars give out too little feedback and the high mass stars are too rare.  An IMF dominated by massive stars could reverse that conclusion  \citep[see also ][]{Wada02}. 

Next, the metal enrichment of AGN/quasar disks and the surrounding inner galaxy \citep{Collin99} by the in situ star formation is then more efficient. Indeed, massive stars live short lives and return metal-rich material into their surroundings much quicker than low mass stars. This would be especially important for a rapid enrichment of young quasar accretion disks. In addition, massive stellar remnants, including stellar mass black holes, will be the end result of stellar evolution of the most massive stars \citep{Heger02}. These black holes may migrate through the accretion disk \citep[e.g.][]{Clarke91} and become interesting sources of gravitational radiation if they travel close enough to the SMBH \citep{Levin03b}.

\subsection{Implications for theories of star formation}

The IMF is a crucial test for any theory of star formation. The IMF found in nearby sites of active or recent star formation, such as Orion Nebula Cluster, appears to be broadly consistent with that of the IMF of the field stars in the solar neighborhood \citep{Hillenbrand97,Meyer00}. Large pe-collapse gas densities or very high rates of star formation in immediate vicinity of galactic centres or in starburst galaxies lead to theoretical suggestions \citep[e.g.,][]{Morris93} or observational claims of a low-mass cutoff in the IMF. None appear to be robust enough, however, as theoretical estimates are widely uncertain and observations may usually be reconciled with the galactic IMF \citep[e.g.,see further references in ][]{Meyer05,Portegies-Zwart02}. 

The suppression of the low mass end of the IMF in \sgra\ star cluster implies that the IMF is indeed not universal, and may significantly differ, at least in extreme conditions that exist near SMBHs in galactic centers. Observations of \sgra\ young star cluster should be used to test theories of star formation in general. It may be that, when taken to extreme conditions of the GC, some of the theories will be inadequate to explain the observed massive stars. Our preliminary analysis along these lines favors theories where the massive stars gain most of their mass through long-range mass accumulation phase \citep[e.g.,][]{Bonnell98,Bonnell01}, such as gas accretion or stellar mergers, rather than a coherent collapse of a sufficiently large reservoir of gas from which the star is then grown \citep[e.g.,][]{McKee03}. We shall present this analysis elsewhere. On the other hand, demonstration that some of the processes may be more important than others in extreme conditions does not necessarily imply that this will be also so for the more "normal" galactic conditions.

\section{Conclusions}

Knowledge of the IMF is important for constraining theories of star formation. 
Young low mass stars cannot be easily observed in the near infrared on the background of high mass stars that radiate $L\sim 10^4 - 10^6 \lsun$ bolometrically. The results of the COUP project \citep[e.g.][]{Feigelson05a,Preibisch05,Flaccomio05,Kastner05} demonstrate that during their PMS of stellar evolution low mass stars are unusually bright X-ray emitters, with a typical luminosity of $L_X \sim 10^{30}$ erg/sec per star. Since \sgra\ region is rather dim in X-rays \citep{Baganoff03a,Muno03,Muno04}, we were able to place an upper limit of $\sim$ few thousand of low mass PMS stars populating the same region as the He-I stars of the better defined stellar ring \citep{Levin03,Genzel03a,Paumard05}. This shows that  formation of  low mass stars should have been  suppressed by a factor of 10 or more relative to the normal galactic IMF in these disks. We also argued that if the young stars were formed in a massive star cluster outside the central parsec then the young low mass stars would have been even more numerous and more visible, and that this scenario of formation of young stars in \sgra\ cluster is therefore ruled out. 

We acknowledge useful discussions with Mike Muno. This paper has been written while SN was attending the program "Jets and Disks" at the Kavli Institute for Theoretical Physics, Santa Barbara.

\bibliographystyle{mnras} 

\end{document}